# Thermodynamics of Charge Regulation Near Surface Neutrality

*Tal Obstbaum and Uri Sivan\**

Department of Physics and the Russell Berrie Nanotechnology Institute, Technion − Israel Institute of Technology, Haifa 3200003, Israel

**ABSTRACT:** The interaction between two adjacent charged surfaces immersed in aqueous solution is known to be affected by charge regulation - the modulation of surface charge as two charged surfaces approach each other. This phenomenon is particularly important near surface neutrality where the stability of objects such as colloids or biomolecules is jeopardized. Focusing on this ubiquitous case, we elucidate the underlying thermodynamics and show that charge regulation is governed in this case by surface entropy. We derive explicit expressions for charge regulation and formulate a new universal limiting law for the free energy of ion adsorption to the surfaces. The latter turns out to be proportional to $k_B T$, and independent of the association energy of ions to surface groups. These new results are applied to the analysis of unipolar as well as amphoteric surfaces such as oxides near their point of zero charge or proteins near their isoelectric point.

*1. Introduction*

Diverse biological and industrial surfaces acquire charge in contact with water by ionization of surface groups. The exposed surface charges attract counterions and repel co-ions to create an electrical double layer comprising an inner layer of bound counterions and a diffuse layer – a cloud of ions distributed according to the self-consistent local electrostatic potential. While the concentration of ions in the diffuse layer is faithfully described by the Poisson-Boltzmann (PB) equation, the density of adsorbed ions in the inner layer depends on multiple factors, including the local electrostatic potential, chemical interactions between ions and surface groups, and hydration of surface groups and counterions. Chemistry and electrostatics are hence entangled in this layer.

When two charged surfaces approach each other, the electrostatic potential induced by each surface modifies the inner layer charge of the other surface in a process called charge regulation. The latter plays a pivotal role in diverse phenomena where modulation of surface charge due to interlayer interaction modulates the force acting between two bodies[1,2]. For example, in cases where the two surfaces are charged with the same polarity, e.g., colloidal suspensions, charge regulation acts to diminish surface charge, resulting in suppressed inter-body repulsion that may in certain cases lead to aggregation[3–8].

Charge regulation has been studied extensively for different models of surface chemistry, leading to approximate expressions or numerical results describing the way surface charge and force vary as a function of surface separation[6,9–25]. With the exception of ref [26], most publications focused on analysis of experimental data and specific models. Few have been dedicated to the underlying thermodynamics[26] and derivation of general rules, which are the subjects of the present study.

To do that, we take advantage of the fact that for common electrolyte concentrations and nearly neutral pH, the overall charge density of the dressed surface (surface plus inner layer) is small compared with the density of surface ion adsorption sites. The net surface charge of many oxides and proteins is in the range of 1-2% of the ionizable surface groups at pH 5.6[27,28] (pH of water in equilibrium with ambient air) and so is their excess protonation level near their point of zero charge (isoelectric point for proteins). This fact follows directly from the definition of the boundary between the inner and diffuse layers. Ions in the diffuse layer move freely and are hence subjected (by definition) to electric potentials on the order or smaller than the thermal energy, $k_B T$. That potential, translated using the Grahame equation, Eq. (A13), to a dressed surface charge, gives for 1mM and 100mM monovalent salt concentrations surface charges of $\approx 0.01 e \cdot nm^{-2}$ and $\approx 0.1 e \cdot nm^{-2}$, respectively. These densities are indeed minute compared with a typical concentration of surface binding sites, e.g., 5 $nm^{-2}$, for silica[28,29].

We show below that under these conditions charge regulation is dominated by surface entropy with enthalpy playing a minor role, if any. Charge regulation in this ubiquitous case is hence universal in the sense of being independent of ion adsorption energies to the surface, leading to insightful, simple expressions for charge regulation. To explore the underlying thermodynamics, we calculate a less studied quantity - the free energy of ion adsorption to the surface. This quantity renders charge regulation an intuitive thermodynamic interpretation, relating it to surface charge compressibility. For the case where the binding sites are either almost occupied or almost empty, we arrive at a new unexpected limiting law

$$\frac{dG}{d\Gamma} \xrightarrow{\sigma \to 0} k_B T. \qquad (1)$$

Here, $G$, $\Gamma$, and $\sigma$ stand for the free energy per unit area, areal density of ionized surface groups and equilibrium surface charge, respectively. $\sigma$ approaches zero near full occupation of a unipolar surface ($\Gamma \to \Gamma_0$ or $\Gamma \to 0$ where $\Gamma_0$ is the areal density of ionizable surface groups).

## 2. Theory

*2.1. Mathematical Description of Charge Regulation*

The surface charge and potential are obtained by solution of the PB equation with the following boundary condition

$$\sigma_d(\psi_s; L) = \sigma_c(\psi_s), \qquad (2)$$

where $\sigma_d$, $\psi_s$ and $L$ are the smeared-out total interfacial charge density according to the Grahame equation, the surface potential corresponding to $\sigma_d$, and the separation between two parallel infinite planar surfaces. $\sigma_c$ is an isotherm denoting the surface charge density for a given chemistry and surface potential. $\sigma$ is the solution of Eq. (2) for specific $L$.

To characterize charge regulation, we define a regulation parameter

$$p \equiv \frac{d\psi_s}{dL} \bigg/ \left(\frac{\partial \psi_s}{\partial L}\right)_\sigma, \qquad (3)$$

the ratio between the actual change of surface potential with $L$ and the same change in the constant charge case. Using Eq. (2) we show in the appendix that $p$ can be expressed in terms of two capacitances

$$p = \frac{c_d}{c_d + c_i}, \qquad (4)$$

with $c_d(\psi_s, L) \equiv (\partial \sigma_d / \partial \psi_s)_L$, and

$$c_i(\psi_s) \equiv -\frac{d\sigma_c}{d\psi_s}. \qquad (5)$$

The first capacitance coincides with the conventional diffuse layer capacitance while the second depicts the inner layer capacitance.

As seen from Eqs. (4) and (5), $c_i$ diverges and $p$ vanishes for a constant surface potential while for constant surface charge $c_i$ vanishes and $p = 1$. In the special case $\psi_s = \sigma = 0$ (e.g., at pzc), both remain zero at all distances regardless of $p$. Note that $p > 1$ implies thermodynamic instability but $p < 0$ is theoretically possible[13].

A simpler version of $p$ was originally introduced by Carnie and Chan (they used $\Delta = 1 - 2p$ )[19]. In that work they formulated a linear charge regulation model by linearizing Eq. (2) around the equilibrium point of isolated parallel plates. This approach is known as the constant regulation approximation[12]. Eq. (4) reduces to their definition in the limit $L \to \infty$. In the general case discussed here, $p$ varies with $L$.

As shown in the appendix, the regulation parameter can also be expressed in terms of surface charge derivatives

$$1 - p = \frac{d\sigma}{dL} \bigg/ \left( \frac{\partial \sigma_d}{\partial L} \right)_{\psi_s}. \tag{6}$$

Eq. (6) together with Eq. (3) lead to a neat relation between surface charge and surface potential regulation

$$\frac{d\psi_s}{dL} \bigg/ \left( \frac{\partial \psi_s}{\partial L} \right)_{\sigma} + \frac{d\sigma}{dL} \bigg/ \left( \frac{\partial \sigma_d}{\partial L} \right)_{\psi_s} = 1. \tag{7}$$

Furthermore, for an arbitrary position $x$ between the two surfaces

$$\frac{d\psi_x}{dL} = p \left( \frac{\partial \psi_x}{\partial L} \right)_{\sigma} + (1-p) \left( \frac{\partial \psi_x}{\partial L} \right)_{\psi_s}. \tag{8}$$

$p$ thus weigh the two limiting cases, constant surface charge (no regulation) and constant surface potential (maximal regulation) to yield the actual potential variation with surface separation. Note that Eq. (8) is exact and not just a linear interpolation formula. For known $p(L)$ the full solution can thus be calculated from the two limiting cases, constant charge, and constant potential. This point has been overlooked in a recent publication that associated such decomposition exclusively with linearized charge regulation models[30]. Furthermore, as indicated by Eqs. (3) and (6), a single limiting case, supplemented with $p(L)$, suffices for calculating the full solution.

2.2. Inner Layer Capacitance

To elucidate the physical origin of charge regulation we express $c_i$ in terms of thermodynamic derivatives. The free energy per unit area of a charged flat surface, separated a distance $L$ from another surface, reads[31]

$$G_1(\sigma, L) = \int_0^\sigma \psi_d(\sigma'; L) d\sigma' + \bar{G}, \tag{9}$$

where $\psi_d$ is the electrostatic potential at the interface between the inner and diffuse layers given by PB equation. The integral yields the electrostatic charging energy while the chemical component, $\bar{G}$, encapsulates the complex electrochemical interactions and internal degrees of freedom in the inner layer.

The second derivative of the free energy per surface with respect to surface charge equals[26,31]

$$\left(\frac{\partial^2 G_1}{\partial \sigma^2}\right)_L = \frac{1}{c_d} + \frac{1}{c_i}. \tag{10}$$

Differentiating Eq. (9) twice with respect to $\sigma$ and substituting the result into Eq. (10), one finds that $c_d^{-1}$ drops and

$$c_i^{-1} = \frac{d^2 \bar{G}}{d\sigma^2} = \sum_{i,j} \frac{\partial \bar{G}}{\partial \Gamma_i \partial \Gamma_j} \frac{d\Gamma_i}{d\sigma} \frac{d\Gamma_j}{d\sigma}, \tag{11}$$

where $(d\Gamma_i/d\sigma)^{-1} = e\sum_m z_m (d\Gamma_m/d\Gamma_i)$. Here, $e$ is the elementary charge, $z_m$ is the valency of the $m$'th species, and $\Gamma_i$ is the surface density of adsorbed ions of the $i$'th species. The total derivatives should be calculated in thermal equilibrium. Eq. (11) relates $c_i$, and hence charge regulation, to the inner-layer charge compressibility.

To elucidate the implications of Eq. (11) we start with Langmuir's model for a single species of monovalent adsorbents. Such a model may describe, for instance, cleaved mica in contact with potassium chloride solution[32]. In this case

$$\bar{G} = G_0 + \epsilon \Gamma + k_B T \Gamma_0 \left[\gamma \ln(\gamma) + (1-\gamma)\ln(1-\gamma)\right], \tag{12}$$

where $G_0$ is some constant, $\epsilon$ is the binding energy of a counterion to its binding site, and $\gamma \equiv \Gamma/\Gamma_0$. The second derivative of the first two terms on the right-hand side vanishes, proving that charge regulation in the Langmuir model is strictly entropic and governed by the number of available arrangements of adsorbents. Substituting Eq. (12) into Eq. (11) one arrives at

$$c_i = \frac{e^2 \Gamma_0}{k_B T} \gamma (1-\gamma). \tag{13}$$

Nearly neutral unipolar surfaces are found when surface binding sites are either nearly full ($\gamma \approx 1$) or nearly empty ($\gamma \approx 0$). In either case $c_i$ is small, indicating incompressible ion adsorption to the surface. The incompressibility reflects divergence of the entropic contribution to the chemical potential, taking place when all binding sites are either occupied or empty. Correspondingly, for finite $L$ the charge regulation parameter, $p$, approaches unity (constant charge limit). Maximal regulation is obtained for $\gamma = 1/2$. An early work mentioned this behavior based on numerical calculations of amphoteric surfaces[17]. Similar results were also shown in a recent work[14].

*2.3. Free Energy of Ion Adsorption*

We turn now to derive a general expression for the free energy change upon adsorption of an extra ion to the surface. At equilibrium, $(\partial G/\partial \sigma)_L = 0$ so $dG/dL = (\partial G/\partial L)_\sigma = -\Pi$, where $G$ is the total free energy for the two-plate system and $\Pi$ is the excess pressure exerted on the surfaces. Eq. (6) then gives

$$\frac{dG}{d\sigma} = \frac{dG}{dL}\frac{dL}{d\sigma} = -\frac{\Pi}{1-p}\left(\frac{\partial \sigma_d}{\partial L}\right)_{\psi_s}^{-1}. \tag{14}$$

Eq. (14) is exact but complicated. Being interested in nearly neutral surfaces we limit ourselves to low potentials, $\phi_s \equiv e\psi_s/k_B T \ll 1$ where $\Pi/(\partial \sigma_d/\partial L)_{\psi_s} = \psi_s$ (Eq. (A17)), thus

$$\frac{dG}{d\sigma} = -\psi_s - \frac{\sigma}{c_i}. \tag{15}$$

The first term on the right-hand side corresponds to the double layer formation energy at constant potential[31,33] (maximal charge regulation) while the second term corrects for weaker regulation.

Note that $\Pi$ is calculated for homogeneous charge distribution. Corrections due to charge discreteness have been shown to be insignificant for small $\sigma$ [34,35]. The reason being that $\Pi$ is dominated by osmotic pressure. Due to solution's neutrality the first order of $\Pi$ with respect to $\sigma$ vanishes (increase in cation concentration is compensated in this order by identical decrease in an-ion concentration) and hence to leading order $\Pi \propto \sigma^2$, just as in the uniform surface charge case (see also Eq. (A16) and the preceding discussion). The similarity between the continuum and discrete cases is further supported by experiments that show excellent fit by continuum mean field theory for both colloidal and small AFM tips[36–39] also at short distances. Deviations are only significant for high surface charge and high salt concetration, which are not relevant to the current manuscript [34,35].

Eq. (15) is true for any adsorption model. Specifically, for Langmuir's model with $c_i$ given by Eq. (13)

$$\frac{dG}{d\sigma} = -\psi_s + \frac{k_B T}{e\gamma}, \tag{16}$$

where we substituted $-e\Gamma_0(1-\gamma)$ for $\sigma$. Notably, the free energy of charging is independent of details such as binding energy of counterions to the surface. This peculiarity traces back to the fact that charging/discharging takes place in equilibrium and that charge regulation is second order in $\gamma$.

Near surface neutrality, $\psi_s \approx 0$ and $\gamma \approx 1$, thus for finite $L$ the free energy of adsorption converges to the thermal energy, Eq. (1), irrespective of chemical energies. This remarkable result results from the compensation of the diverging inverse charge compressibility ( $c_i^{-1} \sim \sigma^{-1}$) by $\sigma \to 0$. The proportionality between $c_i$ and $\sigma$ near neutrality holds generally whenever the surface sites are fully occupied or fully empty near neutrality. For the Langmuir model, it is evident from Eq. (13)

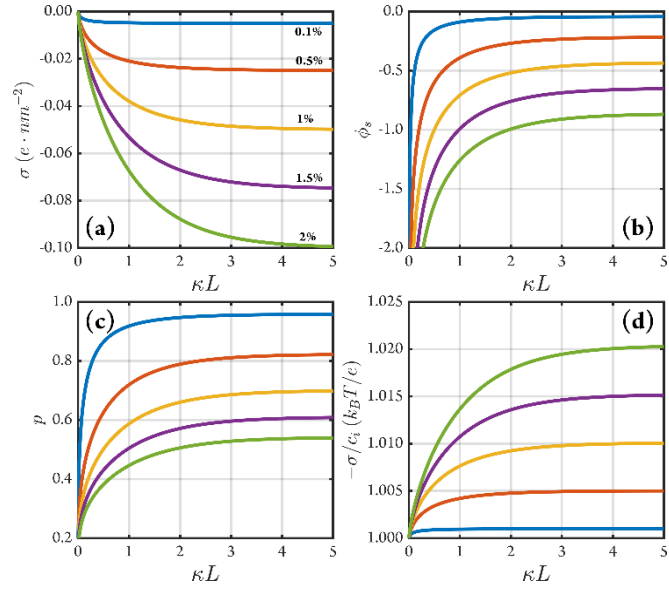

Figure 1. Surface charge (a), normalized surface potential (b), regulation parameter (c), and chemical adsorption free energy (d) for Langmuir's model as a function of separation between surfaces ($\kappa$ is the inverse screening length). Colors refer to different surface charge densities at infinite separation shown in (a) as percentage of the density of ionizable groups, $1-\gamma$.

Eq. (13) together with the linear Grahame equation, Eq. (A13), lead to a simple expression for $p$

$$p^{-1} - 1 = -\phi_s. \tag{17}$$

Substitution of $p$ from Eq. (17) into Eq. (3) leads to an explicit ODE for $\phi_s$

$$\frac{d\phi_s}{dL} = \frac{1}{1-\phi_s}\left(\frac{\partial \phi_s}{\partial L}\right)_\sigma, \tag{18}$$

with the solution

$$\phi_s(L) = -W\left(-\frac{\phi_\infty \exp(-\phi_\infty)}{\tanh(\kappa L/2)}\right). \tag{19}$$

$W$ is the product log (Lambert) function. According to Eq. (19) the surface potential as a function of $L$ depends exclusively on the potential of the isolated surfaces, $\phi_\infty$. Being interested in nearly neutral surfaces we have used here linearized expressions, but the same procedure can be repeated with exact expressions.

Figure 1(a) depicts the surface charge as a function of surface separation, calculated by Eqs. (19) and (A13) for various values of surface charge at infinite separation. Figure 1(b) depicts the corresponding normalized surface potential. Interestingly, to maintain a constant chemical potential as $L \to 0$, the surface potential diverges to compensate the diverging entropy component. Figure 1(c) depicts the regulation parameter. As the two surfaces approach each other, $p$ grows smaller, marking the transition from intermediate regulation at large separations to constant potential near contact. Figure 1(d) depicts the chemical free energy of adsorption given by Eq. (15). The calculations correspond to $\Gamma_0 = 5 nm^{-2}$ and 100mM monovalent salt concentration. As the surface turns neutral with $L \to 0$, the chemical adsorption energy of ions to the surface approaches $k_B T$, independent of surface chemistry.

Generalization of the Langmuir model to several ion species with their distinct adsorption sites is detailed in the appendix.

## 2.4. Amphoteric Surfaces

The case of amphoteric surfaces, such as oxides or proteins, is of broad interest[16–18,40,41]. Such surfaces release protons at high pH, rendering the surface negative. At low pH, they capture excess protons, and the surface turns positive. At a specific intermediate pH called point of zero charge (pzc) or isoelectric point in the case of proteins, the total charge vanishes and system stability against coagulation is compromised. Near that point, charge regulation may play a critical role.

We focus here on oxides but the adaptation to other amphoteric surfaces, such as proteins, is straightforward. Oxides are characterized by hydroxyl surface groups denoted $ROH$, where $R$ represents a surface atom or molecule such as $Si$ in silica. Hydroxyls are amphoteric. They deprotonate at high pH and doubly protonate at low pH[28,42,43]

$$\begin{aligned} RO^- + H^+ &\rightleftarrows ROH \\ ROH + H^+ &\rightleftarrows ROH_2^+ \end{aligned} \quad , \tag{20}$$

with the corresponding association constants $K_1$ for protonation and $K_2$ for double protonation. Accordingly, $pzc = (pK_1 + pK_2)/2$. For most oxides $pK_1$ is a few units larger than $pK_2$, implying that at pzc both deprotonation and double protonation are small. For example, for silica $pK_1 \approx 5$, $pK_2 \approx 0$, and at pzc the fraction of deprotonated and doubly protonated hydroxyls is $1 - \gamma_1 = \gamma_2 \approx 10^{-2.5} \approx 0.3\%$ [28]. Here, $\gamma_1, \gamma_2$ are the fractions of protonated and doubly protonated silanols. Straightforward calculation yields

$$c_i^{pzc} = \frac{2e^2 \Gamma_0 \gamma_2}{k_B T}, \tag{21}$$

where $\Gamma_0$ is the areal density of R groups (Si in the case of silica).

Near pzc $p^{-1} - 1 = 8\pi \Gamma_0 \ell_B \kappa^{-1} \gamma_2 / \tanh(\kappa L/2)$ where we have used Eq. (A13) to calculate $c_d$ and $\ell_B \approx 0.7 nm$ is the Bjerrum length in water at $25°C$. For silica $\Gamma_0 \approx 5 nm^{-2}$ [28,29] and in 100mM monovalent salt, $p \approx [1 + 0.27/\tanh(\kappa L/2)]^{-1}$. For $\kappa L \gg 1$ $p \approx 0.8$ while for $\kappa L \to 0$, $p \to 0$ and regulation is maximal (constant potential limit). Since both $\psi_s$ and $\sigma$ vanish at pzc, Eq. (15) implies $dG/d\sigma = 0$. This situation coincides with the simultaneous constant potential and constant charge scenario discussed earlier.

With the shift of pH away from pzc the system crosses over from the amphoteric case to the unipolar limit. The chemical part of $dG/d\sigma$ varies accordingly from zero at pzc to order $k_B T$ over approximately one pH unit on either side of pzc.

### 3. Discussion

The observation that charge regulation is dominated by entropy remains true also in more sophisticated models. Eq. (11) teaches that in the absence of energy correlations (terms proportional to $\Gamma_i\Gamma_j$) the enthalpic contribution to $c_i$ vanishes and charge regulation is still governed by entropy. In models that take site correlations into account, $c_i$, and hence charge regulation, are affected also by enthalpy[13]. Such models may account for inter-adsorbents interactions such as coulomb, steric, and dispersion forces. However, unlike entropy, enthalpy is analytic in site occupancies and hence contributes at most a constant to $c_i^{-1}$ near either zero or full occupancy. Accordingly, near these points charge regulation is still dominated by entropy.

The effects of finite ion size in the direction perpendicular to the surface and local permittivity changes can be handled within the Stern model, leading to a $\sigma^2$ term in the free energy and a constant contribution to $c_i^{-1}$ (Eq. (11)). Near neutrality, where the entropic contribution diverges this constant can be neglected.

In conclusion, charge regulation near surface neutrality is shown to be dominated by entropy and hence takes a simple form for both unipolar and amphoteric surfaces. In the limit $\sigma \to 0$, the charge regulation parameter of unipolar surfaces approaches unity ($p \to 1$) and the surface charge is pinned to its value at infinite distances. Intricate cancellation of the vanishing inner layer capacitance against the vanishing surface charge leads in this case to a universal limiting law for the free energy of ion adsorption to the surface (one $k_B T$, Eq. (1)). For amphoteric surfaces at pzc, the free energy of ion adsorption vanishes but charge may nevertheless be regulated ($p<1$) due to the finite inner layer capacitance (Eq. (21)). Entropy prevails also away from surface neutrality if energy correlations in site occupancy can be neglected. The analysis presented here elucidates unexplored fundamental aspects of charge regulation that are readily applicable to diverse situations.

## *4. Appendix*

*Regulation Parameter*

To show the equivalence between Eqs. (3) and (4) we take the differential of Eq. (2)

$$\left(\frac{\partial \sigma_d}{\partial \psi_s}\right)_L d\psi_s + \left(\frac{\partial \sigma_d}{\partial L}\right)_{\psi_s} dL = \frac{d\sigma_c}{d\psi_s} d\psi_s. \tag{A1}$$

Rearranging and dividing by $c_d dL$

$$\frac{d\psi_s}{dL}\frac{c_i + c_d}{c_d} = -\left(\frac{\partial \sigma_d}{\partial L}\right)_{\psi_s}\left(\frac{\partial \sigma_d}{\partial \psi_s}\right)_L^{-1}. \tag{A2}$$

From the differential of the surface charge

$$d\sigma = d\sigma_d = \left(\frac{\partial \sigma_d}{\partial \psi_s}\right)_L d\psi_s + \left(\frac{\partial \sigma_d}{\partial L}\right)_{\psi_s} dL, \tag{A3}$$

one obtains the derivative of the surface potential at a constant surface charge

$$\left(\frac{\partial \psi_s}{\partial L}\right)_\sigma = -\left(\frac{\partial \sigma_d}{\partial L}\right)_{\psi_s}\left(\frac{\partial \sigma_d}{\partial \psi_s}\right)_L^{-1}. \tag{A4}$$

The right-hand sides of Eqs. (A2) and (A4) are identical, hence

$$\frac{c_d}{c_d + c_i} = \frac{d\psi_s}{dL}\bigg/\left(\frac{\partial \psi_s}{\partial L}\right)_\sigma, \tag{A5}$$

and

$$p = \frac{c_d}{c_d + c_i}. \tag{A6}$$

To derive Eq. (6) we divide Eq. (A3) by $dL$

$$\frac{d\sigma}{dL} = \left(\frac{\partial \sigma_d}{\partial \psi_s}\right)_L \frac{d\psi_s}{dL} + \left(\frac{\partial \sigma_d}{\partial L}\right)_{\psi_s}, \tag{A7}$$

and substitute Eq. (3) and (A4)

$$1 - p \equiv \frac{d\sigma}{dL}\bigg/\left(\frac{\partial \sigma_d}{\partial L}\right)_{\psi_s}. \tag{A8}$$

Eq. (8) is derived by considering the two equations

$$\frac{d\sigma}{dL} = \left(\frac{\partial \sigma}{\partial L}\right)_{\psi_x} + \left(\frac{\partial \sigma}{\partial \psi_x}\right)_L \frac{d\psi_x}{dL} \qquad (A9)$$

and

$$\left(\frac{\partial \sigma}{\partial L}\right)_{\psi_s} = \left(\frac{\partial \sigma}{\partial L}\right)_{\psi_x} + \left(\frac{\partial \sigma}{\partial \psi_x}\right)_L \left(\frac{\partial \psi_x}{\partial L}\right)_{\psi_s}. \qquad (A10)$$

Substitution of Eqs. (A9) and (A10) into Eq. (A8) yields Eq. (8).

*Free Energy of Adsorption*

To derive Eq. (15) we begin from the one-dimensional linearized PB equation

$$\frac{d^2\psi}{dx^2} = \kappa^2 \psi, \qquad (A11)$$

where $\kappa^2 = \dfrac{e^2}{\varepsilon k_B T}\sum_i n_i^\infty z_i^2$ is the inverse squared screening length calculated for bulk salt concentrations $n_i^\infty$ of valences $z_i$, where $i$ stands for the ionic species. $\varepsilon$ is the medium's permittivity. The solution of Eq. (A11) for symmetrically charged flat planes located at $x = \pm L/2$ is given by

$$\psi(x) = \psi_s \frac{\cosh(\kappa x)}{\cosh(\kappa L/2)}. \qquad (A12)$$

Accordingly, the Grahame equation gives

$$\sigma_d = \varepsilon \kappa \psi_s \tanh(\kappa L/2), \qquad (A13)$$

and

$$\left(\frac{\partial \sigma_d}{\partial L}\right)_{\psi_s} = \frac{\varepsilon \kappa^2 \psi_s}{2\cosh^2(\kappa L/2)}. \qquad (A14)$$

The excess pressure is determined by the osmotic pressure at midplane

$$\Pi = k_B T \sum_i n_i^\infty \left[\exp\left(\frac{e z_i \psi_m}{k_B T}\right) - 1\right], \qquad (A15)$$

where $\psi_m$ is the potential at midplane ($x=0$). The zero order terms in the potential cancel trivially, and the first order terms cancel due to the solution's neutrality. To second order

$$\Pi = \frac{e^2 \psi_m^2}{2 k_B T} \sum_i n_i^\infty z_i^2 = \frac{\varepsilon \kappa^2 \psi_m^2}{2} = \frac{\varepsilon \kappa^2 \psi_s^2}{2 \cosh^2(\kappa L/2)}, \quad (A16)$$

and hence,

$$\Pi \bigg/ \left(\frac{\partial \sigma_d}{\partial L}\right)_{\psi_s} = \psi_s. \quad (A17)$$

Eq. (14) takes accordingly the linear form

$$\frac{dG}{d\sigma} = -\frac{\psi_s}{1-p} = -\psi_s - \frac{\sigma}{c_i}. \quad (A18)$$

*Generalization of Langmuir's model to multiple species*

Generalization of the Langmuir model to several ion species with their distinct adsorption sites gives for the inner layer capacitances

$$\begin{aligned} c_i &= \sum_m c_i^m \\ c_i^m &\equiv \frac{e^2 z_m^2 \Gamma_{0m}}{k_B T} \gamma_m (1-\gamma_m) \end{aligned}, \quad (A19)$$

where $m$ marks the species, $\Gamma_{0m}$ replaces $\Gamma_0$ for the $m$'th species, and $\gamma_m \equiv \Gamma_m/\Gamma_{0m}$. The charge compressibilities of the different species with their respective binding sites thus add. Equilibrium for monovalent ions imposes for any two species $m,n$ with distinct adsorption sites

$$\frac{\gamma_m}{1-\gamma_m} \frac{1-\gamma_n}{\gamma_n} = \frac{K_m a_m}{K_n a_n}, \quad (A20)$$

where $K_m = \exp(-\epsilon_m/k_B T)$ and $a_m$ stand for the association constant and activity of the $m$'th species, respectively. Choosing $K_1 a_1 = \min_m K_m a_m$ one finds

$$c_i^m = c_i^1 \frac{\Gamma_{0m}}{\Gamma_{01}} \frac{K_m a_m}{K_1 a_1} \left(1 + \gamma_1 \left(\frac{K_m a_m}{K_1 a_1} - 1\right)\right)^{-2}. \tag{A21}$$

When $\Gamma_1 \to \Gamma_{01}$ also $c_1 \to 0$, and as a result all $c_m \to 0$. Correspondingly, the system approaches its constant charge limit and $dG/d\sigma \to k_B T/e$.

AUTHOR INFORMATION

**Corresponding Author**

*phsivan@technion.ac.il

**Funding Sources**

This research was supported by the Israeli Science Foundation through grant number 547/17.

**Notes**

The authors declare no competing financial interest.

**Acknowledgements:**

We are grateful to Yariv Kafri for insightful discussions.